\documentclass{PoS}

\usepackage{amssymb}
\usepackage{bm}
\usepackage{amsmath}
\usepackage{newtxtext,newtxmath}
\usepackage[square,comma,numbers,sort&compress]{natbib}
\newcommand{\nn}{\nonumber}

\newcommand{\beq}{\begin{equation}}
\newcommand{\eeq}{\end{equation}}
\newcommand{\beqa}{\begin{eqnarray}}
\newcommand{\eeqa}{\end{eqnarray}}
\newcommand{\bd}[1]{ \mbox{\boldmath $#1$}}

\title{ The angular momentum decomposition in the scalar diquark model}

\ShortTitle{ The angular momentum decomposition in the scalar diquark model}

\author{\speaker{D. A. Amor-Quiroz}\\
        CPHT, CNRS, Ecole polytechnique, IP Paris, F-91128 Palaiseau, France\\
        E-mail: \email{arturo.amor-quiroz@polytechnique.edu}}

\author{M. Burkardt\\
       Department of Physics, New Mexico State University, Las Cruces, NM 88003-0001, U.S.A.\\
        E-mail: \email{burkardt@nmsu.edu}}

\author{C. Lorc\'e\\
        CPHT, CNRS, Ecole polytechnique, IP Paris, F-91128 Palaiseau, France\\
        E-mail: \email{cedric.lorce@polytechnique.edu}}

\abstract{One of the challenges of hadronic physics is to fully understand the structure of the proton. In
particular, there is nowadays a great interest in the decomposition of its total
angular momentum into orbital angular momentum and intrinsic
spin, as well as identifying contributions from valence quarks, sea quarks and gluons.

The most common decompositions of angular momentum are the Jaffe-Manohar (canonical) and Ji (kinetic)
decompositions, which differ in the way contributions are attributed to quarks and gluons.
Using perturbation theory, explicit one-loop calculations found that the difference between
such decompositions vanishes. We justify within the diquark model in QED that the difference
appears at two-loop level, supporting the interpretation of such a difference as originating
from the torque exerted by the spectator system on the struck quark.}

\FullConference{XXVII International Workshop on Deep-Inelastic Scattering and Related Subjects - DIS2019\\
		8-12 April, 2019\\
		Torino, Italy}

\begin{document}

\section{Introduction}
\label{Intro}

The importance of fully understanding the decomposition of the total angular momentum (AM) of the nucleons into orbital angular momentum (OAM) and intrinsic spin contributions has a historical foundation on the so-called \textit{Proton spin crisis}. This term originally refers to the EMC data for deep inelastic muon-proton scattering using a longitudinally polarized target which seemed to imply that the total spin carried by the quarks in the polarized proton amounted only to
1/8 of the spin of the proton~\cite{EMCcrisis1,EMCcrisis2,leader1988crisis}.
Later experiments by COMPASS and HERMES corrected this statement by measuring an approximated 1/3 contribution to the nucleon total AM coming from the intrinsic spin of the valence quarks~\cite{COMPASScrisis,HERMEScrisis}. Introducing the gluon intrinsic spin is still not enough to account the total spin of the nucleon~\cite{DSSVcrisis,NNPDFcrisis}. Therefore, the missing contribution should come from OAM of quarks and gluons.

From a theoretical point of view, the definition of how the nucleon AM can be separated as the sum of the AM of its components is intrinsically ambiguous due to quark-gluon couplings. The most common decompositions of the nucleon AM are the Jaffe-Manohar (canonical)~\cite{JMDecomp} and Ji (kinetic)~\cite{JiDecomp} decompositions. 

For a clear explanation of the differences and the physical interpretation of several OAM decompositions, the reader is referred to the reviews by E. Leader and C. Lorc\'e~\cite{AMcontroversy} and Wakamatsu~\cite{Wakamatsu_review}.
It is sufficient for the moment, to mention that an exploratory lattice calculation of both Ji and Jaffe-Manohar (JM) decompositions of quark OAM was given in 2017 and demonstrates that their difference can be clearly resolved~\cite{EngelhardtLattice}.

Using perturbation theory, explicit one-loop calculations found that the difference between such decompositions vanishes~\cite{JiOneLoop}. The aim of the present document is to show within the scalar diquark model that the difference between both decompositions appears at two-loop level, supporting the interpretation of such a difference as originating from the torque exerted by the spectator system on the struck quark.

\section{Discussion}
\label{Discussion}

At the moment, the only practical way to extract partonic total AM is to use Ji relation
between the quark/gluon total kinetic AM and twist-2
generalized parton distributions (GPDs)~\cite{JiDecomp}
\beqa
J_{a} = \frac{1}{2} \int dx ~ x[H^{a}(x,0,0)+E^{a}(x,0,0)] ~.
\eeqa

The GPD $H^a (x,0,0;Q^2)$ corresponds to the collinear parton distribution function (PDF) $f^a_1(x;Q^2)$, which gives the probability of finding at the scale $Q^2$ a parton of the type $a$ with a fraction $x$ of the longitudinal momentum of the parent nucleon.
The GPD $E^a (x,0,0;Q^2)$ is related to the distortion of PDFs in impact parameter space if the target is  not  a  helicity  eigenstate,  but  has  some  transverse  polarization~\cite{burkardtGPDS}.

As well as GPDs, transverse-momentum distributions (TMDs) are expected to provide some information about the OAM content of the nucleon. Not only because many TMDs would be identically zero in absence of parton OAM, but also because they introduce some constraints to the wave function of the nucleon~\cite{CedricOAM}. 

Unfortunately, it is nowadays unclear whether a general relation between TMDs and OAM can be derived from first principles. Nevertheless, Burkardt proposed that there may be a close connection between the GPD $E$ and the Sivers TMD $f^{\perp}_{1T}$, given by the following expression
\beqa
\int d^2 k_\perp \frac{\bd{k}^2_\perp}{2 M^2}
f^{\perp}_{1T}(x,\bd{k}^2_\perp)
& \propto &
\int d^2 b_\perp \mathcal{I}(\bd{b}_\perp) 
(\bd{S}_T \times \bd{\partial}_{b_\perp})_z
\mathcal{E}^{q}(x,\bd{b}^2_\perp)
\nn\\
\mathcal{E}^{q}(x,\bd{b}^2_\perp)
& = &
\int  \frac{d^2 \Delta_\perp}{(2\pi)^2}
e^{-i\bd{b}_{\perp} \cdot \bd{\Delta}_\perp}
E(x,0,-\bd{\Delta}^2_\perp) ~.
\label{Lensing}
\eeqa
The lensing function $\mathcal{I}(\bd{b}_\perp)$ accounts for the effect of attractive initial/final state interactions (ISI/FSI) due to soft gluon rescattering in deep-inelastic  and  other  high-energy scattering. They provide some transverse momentum to the outgoing quark, deflecting the latter towards the center of the target~\cite{BurkardtLensing0}. 

Both ISI and FSI can effectively be taken into account by introducing an appropriate Wilson line phase factor in the definition of the distribution functions of quarks in the nucleon.
It is therefore postulated that the chromodynamic lensing mechanism can be related to the Wilson phase contribution and furthermore, to the torque generated on the struck quark with respect to the spectator system~\cite{BurkardtLensing1}. 

Even if the relation in eq.~(\ref{Lensing}) is intuitive and is supported by some model calculations~\cite{BurkardtLensing1,BurkardtAndHwang,BacchetaLensing,EikonalLensing}, we emphasize that so far no model-independent relations between GPDs and TMDs have been established. In the case of the present calculation we do not require the possibility of factorizing $f^{\perp}_{1T}$ into a distortion effect (described by the GPD $E^q$)  times a FSI.
We argue, nevertheless, that a non-vanishing Sivers function is a necessary condition for a non-vanishing difference between Ji and JM decompositions, as both mechanisms have the same physical foundations in ISI/FSI.

In terms of the mathematical formulation, Ji decomposition is related to the usual gauge covariant derivative $D_\mu$ while JM decomposition is associated with the \textit{pure} gauge covariant derivative $D^{pure}$ defined as $D^{pure}_\mu=\partial_\mu-ie_q A^{pure}_\mu$, where the notation refers to the \textit{Chen et al.}~\cite{ChenEtAl} splitting of the fields provided by~\cite{Hatta2011,Lorce2012}
\beqa
    A_\mu &=& A^{pure}_\mu+A^{phys}_\mu 
    \nn\\
    F^{pure}_{\mu\nu} &=&
    \partial_\mu A^{pure}_{\nu}
    -\partial_\nu A^{pure}_{\mu}
    +ig[A^{pure}_{\mu},A^{pure}_{\nu}]
    = 0 ~.
\eeqa

The \textit{physical} part of the field ${A}^{phys}_\mu$ can be fixed in any gauge. In the present document we set the light-cone gauge $A^{+}_{phys}=0$ since it results convenient for simplifying some calculations, as the pure derivative coincides with the common partial derivative provided by $\partial_\mu$ in such gauge. Besides, FSI can be taken into account with a gauge link involving the gauge potential at the spatial infinity~\cite{JiFSIinGaugeLinks}.
Any function depending on the difference 
$ D_\mu - D^{pure}_\mu $ can therefore be expressed in terms of the physical fields as
\beqa
    \langle \bar{\Psi } \gamma^+  D_\mu \Psi \rangle
    ~ - ~
    \langle \bar{\Psi } \gamma^+  D^{pure}_\mu \Psi \rangle
    ~ = ~ -
    \langle \bar{\Psi } \gamma^+  ie_q A^{phys}_\mu \Psi \rangle ~.
    \label{Observables}
\eeqa

An example of the above is obtained for the perpendicular components of eq.~(\ref{Observables}) after multiplying it by a factor of $-i$ in order to identify the terms on the left-hand side with Ji and JM twist-2 transverse momentum operators respectively as
\beqa
\langle \bd{k}_\perp \rangle_{Ji} -  \langle \bd{k}_\perp \rangle_{JM}
    = -e_q
    \langle \int d^3 r ~ \bar{\Psi }(\bd{r}) \gamma^+ \bd{A}^{phys}_{\perp}(\bd{r}) \Psi (\bd{r}) \rangle ~,
\eeqa
while the right-hand side of eq.~(\ref{Observables}) is denoted as the potential momentum~\cite{WakamatsuPotOAM}. Analogously we define the potential OAM as the difference between Ji and JM OAM in leading twist approximation
\beqa
    \langle  L \rangle_{Ji}
    - \langle  \mathcal{L} \rangle_{JM}
    = - e_q
    \langle \int d^2 r_{\perp} ~ 
    \bar{\Psi }(\bd{r}) \gamma^+  
    (\bd{r}_{\perp} \times \bd{A}^{phys}_{\perp} (\bd{r}))_z 
    \Psi (\bd{r}) \rangle ~.
\eeqa
The potential momentum and potential OAM have the physical interpretation as the change in $\bd{k}_\perp$ and OAM experienced by the struck quark due to the color Lorentz forces as it leaves the target~\cite{BurkardtSivers}. 
For a recent discussion of the potential OAM in the Landau problem, see~\cite{WakamatsuLandau}.\\

We compute both the potential momentum and potential OAM in the framework of a \textit{simple} scalar diquark model (SDM) of the nucleon in the one-loop approximation. The model is based on the assumption that the nucleon splits into a quark and a scalar diquark structure, which are regarded as elementary fields of the theory~\cite{BrodskySDMandFSI}. 

While the active quark interacts with the photon, the diquark acts as the spectator and vice versa. For the sake of simplicity, we consider only abelian couplings of the fields for the ISI/FSI. 
In the present model the target has no electromagnetic charge emulating the fact that  in  QCD  a  hadron  is  color  neutral.  This  condition directly implies a relation between the coupling constants, though they are left as independent parameters in the following.

Despite of its naivet\'e, such a model provides analytic results while giving an estimate of the magnitude of the observables. Additionally, it has the feature of maintaining explicit Lorentz covariance.

\subsection{Potential Momentum}
\label{PotMomentum}

The Sivers mechanism gives rise to a non-zero average transverse momentum of partons inside a transversely polarized target. This is not forbidden by conservation of momentum as long as the contributions from the partons sum up to zero in what is known as
Burkardt sum rule and such a property is irrespective of the decomposition used~\cite{BurkardtSumRule}. 
In the case of the SDM it takes the form     
$ \sum_{a=q,s} \langle \bd{k}^{a}_\perp \rangle  = \bd{0}_{\perp}$ where $q$ and $s$ denote the quark and the scalar diquark, respectively.

By means of explicit two-loop calculations, we found that Burkardt sum rule is trivially fulfilled for Ji decomposition as 
$ \langle \bd{k}^{q}_\perp \rangle_{Ji}  
= \langle \bd{k}^{s}_\perp \rangle_{Ji} 
= \bd{0}_{\perp}  $. 
This agrees with the physical interpretation of Ji decomposition as no ISI/FSI that can provide a spin asymmetry are taken into account. 

It is because of the local description of the full covariant derivative that Ji transverse momentum of the partons is even under naive time reversal transformations and consequently it has to vanish to all orders.
On the contrary, JM decomposition refers to a non-local notion of a derivative and thus contains a mixed symmetry under naive time reversal.
Consequently the potential momentum indirectly determines JM transverse momentum of the quark. Two-loop level calculations already provide some structure to the partons transverse momentum.  
\beqa
    & \langle \bd{k}^i_\perp \rangle_{JM}
    =  e_q
    \langle \int d^3 r ~ \bar{\Psi }(r) \gamma^+ \bd{A}^{i}_{phys}(r) \Psi (r) \rangle 
& \nn\\
    & = \frac{1}{6}
    \left( \frac{g}{4 \pi\epsilon} \right)^2
    \frac{\epsilon^{ij}_{T} s^{j}_\perp }{ (4\pi)^2}(3m_q + M)\pi e_s e_q + \mathcal{O}\left(\frac{1}{\epsilon}\right)~. &
    \label{k_average}
\eeqa
The non-vanishing contribution come from the gauge-link at light-cone infinity.
In the expression above, $m_q$ denotes the mass of the quark, $M$ is the mass of the nucleon, $e_q$ and $e_s$ are the photon-quark and photon-diquark couplings respectively, $g$ is the coupling constant of the point-like scalar quark-nucleon-diquark vertex and $s_\perp$ is the transverse polarization of the target. The factor $4\pi\epsilon$ comes from the dimensional regularization prescription for the integral over transverse momentum, meaning that only the divergent piece of the amplitudes is retained. Moreover, the stability condition for the target state is $M<m_q+m_s$. 

Some analytic expressions for the Sivers function $f^{\perp a}_{1T}$ have already been provided for the SDM~\cite{JiFSIinGaugeLinks,GoekeSiversSDM}. They can be related to the average transverse momentum found in eq.~(\ref{k_average}) using the following identity~\cite{MeissnerSivers}
\beqa
 \langle \bd{k}^{i}_{\perp} \rangle 
 =
 - \int dx \int d^2 k_{\perp} 
 \bd{k}^{i}_{\perp} 
 \frac{\epsilon^{jk}_{T}\bd{k}^{j}_{\perp} \bd{s}^{k}_{\perp}}{M}
 f^{\perp }_{1T}(x,\bd{k}^2_\perp) ~.
\eeqa

The difference between Ji and JM decompositions for the transverse momentum appears at two-loop level. This supports the interpretation of such a difference as originating from the torque exerted by the spectator system on the struck quark.

\subsection{Potential OAM}
\label{PotOAM}

Using the light-cone gauge we also computed Ji and JM OAM in the absence of gauge fields within the SDM. As a consequence, both decompositions coincide as already observed for the same model at the level of impact parameter distributions~\cite{CedricResults}
\beqa
    \langle  L \rangle_{Ji}
    = \langle  \mathcal{L} \rangle_{JM}
    =
    \langle \int d^2 r_{\perp} ~ \bar{\Psi }(\bd{r}) \gamma^+  
    \bd{r}_{\perp} \times i\nabla_{\perp} \Psi (\bd{r}) \rangle ~.
\eeqa
    
This means the potential OAM is zero at one-loop order and any difference between kinetic and canonical decompositions can only be observed at higher order. Anyhow we are able to provide analytical expressions for the OAM of the partons as
\beqa
   \langle \mathcal{L} \rangle_q (x) 
   &=& 
   \frac{ g^2 }{16\pi^3}(1-x)^2 \int d^2 k_{\perp}  \frac{ \bd{k}^2_\perp }{ ( \bd{k}^2_\perp + \Lambda^2_q )^2 } \nn\\
   \langle \mathcal{L} \rangle_s (x) 
   &=&  
   \frac{x}{1-x} \langle \mathcal{L} \rangle_q (x)  ~.
   \label{1-loopOAM}
\eeqa

From the expression in eq.~(\ref{1-loopOAM}) it is easy to check that Ji sum rule~\cite{JiDecomp} is also fulfilled
\beqa
     J(x)= \langle \mathcal{L} \rangle_q (x)+ \langle \mathcal{L} \rangle_s (x) + \frac{1}{2}g^q_1 (x) 
 ~~ \Longrightarrow ~~ J = \int dx J(x) =\frac{1}{2} ~,
\eeqa
where $g^q_1 (x)$ is the helicity PDF that accounts for the contribution from the quark to the total AM.\\

Two-loop calculations for the potential OAM will be published elsewhere. The appearance of Sivers effect at the same order strongly suggest a non-vanishing difference of Ji and JM decompositions.

\section{Summary and Outlook}
\label{Conclusions}
 
The potential momentum was computed for the scalar diquark model as an equivalent to obtaining the difference between Ji and JM decompositions, which turns out to be non-zero in perturbation theory at $\mathcal{O}(g^2 e_q e_s)$.

\vspace*{-0.05cm}
 
As suggested by eq.~(\ref{Lensing}), in order to generate the Sivers effect the  impact  parameter  distribution  of  unpolarized quarks in a transversely polarized target has to be distorted (non-zero $E$) and  the  fragmenting  struck  quark has to experience either initial or final state interactions with the target spectators.
A non-vanishing Sivers function therefore suggests a difference between Ji and JM decompositions to appear at two-loops order due to the presence of initial or final state interactions between the struck quark and the spectator system.

For the moment, only one-loop calculations were carried for Ji and JM decompositions for OAM, where they coincide as expected from the lack of initial or final state interactions.
At this order it was also explicitly verified that Ji sum rule is satisfied.



\begin{thebibliography}{89}

\bibitem{EMCcrisis1} J. J. Aubert \textit{et al.} [European Muon Collaboration], Nuc. Phys. B \textbf{259}, (1985) 189.

\bibitem{EMCcrisis2} J. Ashman \textit{et al.} [European Muon Collaboration], Phys.Lett. B \textbf{206}, (1988) 364.

\bibitem{leader1988crisis} E. Leader and M. Anselmino, Z. Phys. C \textbf{41}, (1988) 239.

\bibitem{COMPASScrisis} V. Y. Alexakhin \textit{et al.} [COMPASS Collaboration], Phys.Lett. B \textbf{647}, (2007) 8.

\bibitem{HERMEScrisis} A. Airapetian \textit{et al.} [HERMES Collaboration], Phys.Rev. D \textbf{75}, (2007) 012007.

\bibitem{DSSVcrisis} D. de Florian et al [DSSV Collaboration], Phys Rev. Lett. \textbf{113}, (2014) 012001.

\bibitem{NNPDFcrisis} E. R. Nocera \textit{et al.} [NNPDF Collaboration], Nuc. Phys. B \textbf{887}, (2014) 276.

\bibitem{JMDecomp} R. L. Jaffe and A. Manohar, Nucl. Phys. B \textbf{337}, (1990) 509.

\bibitem{JiDecomp} X. Ji, Phys. Rev. Lett. \textbf{78},  (1997) 610.

\bibitem{AMcontroversy} E. Leader and C. Lorc\'e, Phys. Rep. \textbf{541.3}, (2014) 163.

\bibitem{Wakamatsu_review} M. Wakamatsu, Int. J. Mod. Phys. A \textbf{29}, (2014) 1430012.

\bibitem{EngelhardtLattice} M. Engelhardt, Phys. Rev. D \textbf{95}, (2017) 094505.

\bibitem{JiOneLoop} X. Ji \textit{et al.}, Phys. Rev. D \textbf{93}, (2016) 054013.

\bibitem{burkardtGPDS}  M. Burkardt,  Int. J. Mod. Phys. A, \textbf{18}(02), (2003) 173.

\bibitem{CedricOAM} K. F. Liu and C. Lorc\'e, Eur. Phys. J. A \textbf{52}(6), (2016) 160.

\bibitem{BurkardtLensing0} M. Burkardt, Phys. Rev. D \textbf{66}, (2002) 114005.

\bibitem{BurkardtLensing1} M. Burkardt, Nucl. Phys. A \textbf{735}, (2004) 185.

\bibitem{BurkardtAndHwang} M. Burkardt and D. S. Hwang, Phys. Rev. D \textbf{69}, (2004) 074032.

\bibitem{BacchetaLensing} A. Bacchetta and M. Radici, Phys. Rev. Lett. \textbf{107}, (2011)
212001.

\bibitem{EikonalLensing} L. Gamberg and M. Schlegel, Phys. Lett. B \textbf{685}, (2010) 95.

\bibitem{ChenEtAl} X.S. Chen \textit{et al.}, Phys. Rev. Lett. \textbf{100}, (2008) 232002.

\bibitem{Hatta2011} Y. Hatta, Phys. Lett. B \textbf{708}, (2011) 186.

\bibitem{Lorce2012} C. Lorc\'e, Phys. Lett. B \textbf{719}, (2005) 185.

\bibitem{JiFSIinGaugeLinks} X. Ji and F. Yuan, Phys. Lett. B \textbf{543}, (2002) 66.

\bibitem{WakamatsuPotOAM}  M. Wakamatsu, Phys. Rev. D \textbf{81} (2010), 114010.

\bibitem{BurkardtSivers} M. Burkardt, Phys. Rev. D \textbf{88}, (2013) 014014.

\bibitem{WakamatsuLandau}  M. Wakamatsu, Y. Kitadono and P.-M. Zhang,
Annals Phys. \textbf{392} (2018), 287.

\bibitem{BrodskySDMandFSI} S. J. Brodsky, D.S. Hwang and I. Schmidt, Phys. Lett. B \textbf{530}, (2002) 99.

\bibitem{BurkardtSumRule} M. Burkardt, Phys. Rev. D \textbf{69}, (2004) 091501.

\bibitem{GoekeSiversSDM} K. Goeke \textit{et al.}, Phys. Lett. B \textbf{637}, (2006) 241.

\bibitem{MeissnerSivers} S. Meissner, A. Metz, and K. Goeke, Phys. Rev. D \textbf{76}, (2007) 034002.

\bibitem{CedricResults} C. Lorc\'e, L. Mantovani and B. Pasquini, Phys. Lett. B \textbf{776}, (2018) 38.

\end{thebibliography}
\end{document}